\documentclass[12pt]{article}
\usepackage{epsfig}
\newcommand{\mqbar}{\bar{m}_q^2}
\newcommand{\msqbar}{\bar{m}_{\tilde{q}_j}^2}
\newcommand{\mstbar}{\bar{m}_{\tilde{t}_j}^2}
\newcommand{\msbbar}{\bar{m}_{\tilde{b}_j}^2}
\def\MM{{\cal M}}
\def\Veff{{V_{\rm eff}}}

\def\b{\beta}

\def\th{\theta}

\def\barm{{\bar m}}
\def\del{\partial}
\def\half{{1\over2}}
\def\st{{\tilde t}}
\def\sb{{\tilde b}}
\def\sq{{\tilde q}}
\def\absv#1{\left|#1\right|}
\def\expecv#1{\left<#1\right>}

\def\gtsim{\mathrel{\hbox{\raise0.2ex
\hbox{$>$}\kern-0.75em\raise-0.9ex\hbox{$\sim$}}}}
\def\ltsim{\mathrel{\hbox{\raise0.2ex
\hbox{$<$}\kern-0.75em\raise-0.9ex\hbox{$\sim$}}}}
\def\tanb{\tan\beta}
\begin{document}
\begin{titlepage}
\begin{flushright}
SAGA--HE--195\\
KYUSHU--HET--62\\
\date={November~14, 2002}
\end{flushright}
\vspace{24pt}
\centerline{\Large {\bf CP Violation in the Higgs Sector and}}
\vspace{8pt}
\centerline{\Large {\bf Phase Transition in the MSSM}}
\vspace{8pt}
\vspace{24pt}
\begin{center}
Koichi \textsc{Funakubo}$^{a,}$\footnote{E-mail: funakubo@cc.saga-u.ac.jp},
Shuichiro \textsc{Tao}$^{b,}$\footnote{E-mail: tao@higgs.phys.kyushu-u.ac.jp}
and
Fumihiko \textsc{Toyoda}$^{c,}$\footnote{E-mail: ftoyoda@fuk.kindai.ac.jp}
\end{center}
\begin{center}
{\it $^{a)}$Department of Physics, Saga University,
Saga 840-8502 Japan}
\vskip 0.2 cm
{\it $^{b)}$Department of Physics, Kyushu University, 
Fukuoka 812-8581 Japan}
\vskip 0.2 cm
{\it $^{c)}$Kyushu School of Engineering, Kinki University, 
Iizuka 820-8555 Japan}
\end{center}
\baselineskip=20pt
\vskip 1.0 cm
\bigskip\bigskip
\abstract{%
We investigate the electroweak phase transition in the presence of a large
$CP$ violation in the squark sector of the MSSM. When the $CP$ violation is
large, scalar-pseudoscalar mixing of the Higgs bosons occurs and a large $CP$ 
violation in the Higgs sector is induced. It, however, weakens first-order phase
transition before the mixing reaches the maximal. Even when the $CP$ violation
in the squark sector is not so large that the phase transition is strongly 
first order, the phase difference between the broken and symmetric phase regions
grows to $\mathcal{O}(1)$, which leads to successful baryogenesis, when the charged Higgs
bosons is light.
}
%
\end{titlepage}
%
%
%
\section{Introduction}
The baryon asymmetry of the universe (BAU) is one of the most obvious
facts, which has been a longstanding problem in astrophysics\cite{KolbTurner}.
To explain the light-element abundances within the framework of the 
standard big-bang nucleosynthesis, it is required that \cite{PDG}
\begin{equation}\label{bau}
  \frac{n_B}{s}=(0.21-0.90)\times10^{-10}.
\end{equation}
It is well known that in order to obtain this asymmetry starting from a
baryon-symmetric state, three requirements must be satisfied: 
baryon number violation, $C$ and $CP$ violation,
and departure from equilibrium\cite{Sakharov}.
In general, electroweak theories satisfy baryon number violation through chiral anomaly
and have a possibility to generate the BAU\cite{EWBreview}.
In the minimal standard model(MSM), the main source of $CP$ violation comes from 
the phase $\delta_{\rm KM}$ in the Cabibbo-Kobayashi-Maskawa (CKM) quark mixing matrix.
Although this phase is able to account for the experimentally observed $CP$
violation in the neutral $K$-mesons and, as recently measured, in the $B_d$
system, it has been shown that it is not possible to generate 
sufficient BAU through $\delta_{\rm KM}$. 
Furthermore, the strength of the phase transition is so weak in the MSM 
with the Higgs scalar heavier than $115{\rm GeV}$\cite{MSM-limit, MSM-lattice}\  
that the universe is approximately in equilibrium, when baryon number changing
process is in effective.\par
In the context of supersymmetric (SUSY) extension of the MSM, it has
been pointed out that in the presence of a light stop the electroweak
phase transition (EWPT) can be strong enough for baryogenesis to take place\cite{light-stop}. 
Moreover, SUSY models contain many complex parameters as new $CP$-violating sources 
in addition to $\delta_{\rm KM}$; 
the Higgs bilinear term, $\mu$ and
soft SUSY breaking terms (gaugino masses and scalar trilinear couplings)
\cite{MSSM-CPV}.
Besides these complex parameters, the relative phase of the expectation values of 
the two Higgs doublets $\theta$ might be induced by radiative and finite-temperature
effects, although it vanishes at the tree level.
Without any complex parameter, the phase $\theta$ could be induced by loop effects of
SUSY particles. At zero temperature, this idea of spontaneous $CP$ violation was studied in 
the minimal supersymmetric standard model (MSSM) and it is pointed out that there
inevitably appears a pseudoscalar boson as light as several GeV\cite{Maekawa}.
The same mechanism at finite temperatures was suggested in Ref.~\cite{Comelli} and
extensively studied by some of us\cite{trans-CPV}. 
They found that $\theta$ could be large only in the transient region between the symmetric 
and broken phase regions, with a pseudoscalar Higgs whose mass satisfies 
$m_A\ltsim 85\mbox{GeV}$.
This mechanism was appealing in that such a large $\theta$ can produce sufficient BAU
and it does not induce a large $\theta$ in our vacuum, which is consistent with the bound from neutron electric dipole moment (nEDM).
This scenario, however, is now excluded since the mass of the pseudoscalar must satisfy
$m_A\ge 90.1\mbox{GeV}$\cite{PDG}.\par
Nonzero $\theta$ is also induced from complex parameters in the MSSM.
But magnitude of explicit $CP$ violation is constrained by nEDM measurements.
For example, the physical $CP$ phase relevant to the EDM must be as small as 
$\mathcal{O}(10^{-3})$ when masses of the SUSY particles are weak scale, or
they are heavier than $1\mbox{TeV}$ when the phases are 
$\mathcal{O}(1)$\cite{MSSM-CPV, Kizukuri}.
Recently it was observed that the spectrum and interactions of the neutral Higgs bosons are affected by a large explicit $CP$ violation in the third generation of squarks sector,
which is not restricted by the nEDM constraints\cite{Carena}.
The authors found that when the imaginary part of the product of $\mu$ and
the Higgs-squark trilinear coupling is large, the lightest Higgs boson $H_1$, which is
composed of the scalar and pseudoscalar Higgs fields, becomes much lighter than
the present bound $115\mbox{GeV}$, but is hard to be observed, since its couplings
to the gauge boson and to the bottom quarks is very small.\par
One may think that this $CP$ violation can generate sufficient BAU in the MSSM 
for parameter sets allowed by experiments. 
Our purpose is to investigate the effects of the explicit $CP$ violation on
the EWPT in the MSSM and to evaluate the magnitude of $CP$ violation relevant 
to electroweak baryogenesis, which should be measured at the transition temperature.
The organization of the paper is as follows.
In Section~2, we formulate the effective potential of the Higgs fields including the
one-loop corrections from the gauge bosons and the third generation of quarks and squarks,
both at zero and finite temperatures.
The masses of the three neutral Higgs bosons and the charged Higgs boson are defined as
the derivatives of the effective potential at the zero-temperature vacuum.
The mass formulas are almost the same as those in Ref.~\cite{Carena} except for inclusion
of the gauge boson contributions.
In Section~3, we study the EWPT for parameter sets, which are consistent with
the mass bounds on the neutral lightest and charged Higgs bosons, in the absence
of the explicit $CP$ violation. Next we introduce the phase of the trilinear coupling,
examine how the strength of the phase transition changes, and evaluate the magnitude
of the $CP$ violation relevant to electroweak baryogenesis. 
Section~4 is devoted to concluding remarks. 
We summarize the formulas for the Higgs masses in Appendices.
%
%
%
\section{Effective potential of the MSSM}
We consider the MSSM that has the following superpotential,
%
\begin{equation}
\mathcal{W}=\epsilon_{ij}\left( f_{AB}^{(d)}H_d^iQ_A^jD_B
  -f_{AB}^{(u)}H_u^iU_B - \mu H_d^iH_u^j\right).
\label{super-potential}
\end{equation}
Besides supersymmetric lagrangian, the low-energy MSSM contains the soft-SUSY-breaking terms
%
\begin{eqnarray}
\mathcal{L}_{\rm soft} =
&-& \tilde{m}_1^2\Phi_d^\dag\Phi_d - \tilde{m}_2^2\Phi_u^\dag\Phi_u
  +\epsilon_{ij}\left(\tilde{m}_3^2\Phi_d^i\Phi_u^j+{\rm h.c.}\right)
  \nonumber \\
&-& m_{\tilde{q}AB}^2\tilde{q}_{AL}^\dag\tilde{q}_{BL}
  -m_{\tilde{d}AB}^2\tilde{d}_{AR}^\dag\tilde{d}_{BR}
  -m_{\tilde{u}AB}^2\tilde{u}_{AR}^\dag\tilde{u}_{BR}
  \nonumber \\
&-& \epsilon_{ij}\left[
  \left(f^{(d)}A^{(d)}\right)_{AB}\Phi_d^i\tilde{q}_{AL}^j\tilde{d}_{BR}^*
  -\left(f^{(u)}A^{(u)}\right)_{AB}\Phi_u^i\tilde{q}_{AL}^j\tilde{u}_{BR}^*
  +{\rm h.c.}\right].
\label{soft-breaking-term}
\end{eqnarray}
We calculate the effective potential of the Higgs fields by taking into account
the one-loop contributions from the gauge bosons and the third generation of quarks
and squarks. We consider the gauginos to be heavy enough to decouple so that the
most dangerous contribution to nEDM from the gluino is negligible.
The correction from the leptons, the other quark and squarks can be neglected
because of their small Yukawa couplings to the Higgs fields.
Now the effective potential at zero temperature is given by
\begin{equation}
V_{\rm eff}(\Phi_d,\Phi_u;T=0)  = V_0(\Phi_d,\Phi_u) + \Delta_0V(\Phi_d,\Phi_u),
           \label{effective-potential-0}
\end{equation}
where $V_0(\Phi_d,\Phi_u)$ is the tree-level potential
\begin{equation}
V_0 = m_1^2\Phi_d^\dag\Phi_d + m_2^2\Phi_u^\dag\Phi_u
- \left(m_3^2\epsilon_{ij}\Phi_d^i\Phi_u^j + {\rm h.c.}\right)
+ \frac{g_2^2 + g_1^2}{8}\left(\Phi_d^\dag\Phi_d - \Phi_u^\dag\Phi_u
  \right)^2
+ \frac{g_2^2}{2}\left|\Phi_d^\dag\Phi_u\right|^2,
\label{eq:def-V0}
\end{equation}
and $\Delta_0V(\Phi_d,\Phi_u)$ is the one-loop correction written as
\begin{eqnarray}
 \Delta_0V
 &\equiv& \frac{N_C}{32\pi^2}\sum_{q=t,b}
  \left[\sum_{j=1,2}
    \left(\msqbar\right)^2\left(\log\frac{\msqbar}{M^2}-\frac{3}{2}\right)
    -2\left(\mqbar\right)^2\left(\log\frac{\mqbar}{M^2}-\frac{3}{2}\right)
  \right] \nonumber \\
 & & +\frac{3}{64\pi^2}\left[
  \left(\bar{m}_Z^2\right)^2\left(\log\frac{\bar{m}_Z^2}{M^2}-\frac{3}{2}\right)
    +2\left(\bar{m}_W^2\right)^2\left(\log\frac{\bar{m}_W^2}{M^2}
    -\frac{3}{2}\right)\right].
\label{one-loop-correction}
\end{eqnarray}
Here $\mqbar$, $\mstbar$, $\msbbar$ $\bar{m}_Z^2$ and $\bar{m}_W^2$ are 
field-dependent masses defined by (\ref{quark-masses}), (\ref{stop-masses}), 
(\ref{sbottom-masses}), (\ref{mzbar}) and (\ref{mwbar}), respectively
$M$ denotes the renormalization scale, which we choose such that the loop corrections
vanish at the vacuum.
The expression (\ref{one-loop-correction}) is the same as the one-loop correction
in \cite{Carena}, except for our inclusion of the gauge-boson contributions, which
strengthen the first-order EWPT.\par
It is well known that the masses of the Higgs bosons receive large corrections 
from the loops of the top quark and squarks\cite{corr-Higgs-mass}.
Here, the masses of the Higgs bosons are defined by the second derivative of the
effective potential at the vacuum. To evaluate them, we parameterize the Higgs
fields by the vacuum $(v_d, v_u, \theta)$ and fluctuation around it as
\begin{equation}
\Phi_d = 
  \left(\begin{array}{c}\frac{1}{\sqrt{2}}(v_d+h_d+ia_d)\\ \phi_d^-\end{array}\right),
\quad
\Phi_u =
e^{i\theta}
  \left(\begin{array}{c}\phi_u^+\\ \frac{1}{\sqrt{2}}(v_u+h_u+ia_u)\end{array}\right).
\label{higgs-field-parametrization}
\end{equation} 
In the following, we represent the quantities evaluated at the vacuum by 
$\left<\cdot\cdot\right>$, that is, evaluated with all the fluctuations being set to zero.
Requiring that the first derivatives of the effective potential with respect to 
the neutral Higgs fields evaluated at the vacuum vanish, that is,
\begin{equation}
 \expecv{{{\del\Veff}\over{\del h_d}}} = \expecv{{{\del\Veff}\over{\del h_d}}} = 0,
 \qquad
 \expecv{{{\del\Veff}\over{\del a_d}}} = \expecv{{{\del\Veff}\over{\del a_d}}} = 0,
        \label{eq:tadpole-condition}
\end{equation}
we can express $m_1^2$ and $m_2^2$ in (\ref{eq:def-V0}), 
in terms of ${\rm Re}(m_3^2e^{i\theta})$, $\tanb= v_u/v_d$ and the particle masses 
from the first equation, and 
${\rm Im}(m_3^2e^{i\theta})$ in terms of the particle masses from the second equation.
Now the mass-squared matrix of the neutral Higgs scalars is expressed as
\begin{equation}
\left(\begin{array}{cc}
  \mathcal{M}_S^2 & \mathcal{M}_{SP}^2\\
  \left(\mathcal{M}_{SP}^2\right)^T & \mathcal{M}_P^2
\end{array}\right),       \label{eq:mass-matrix-1}
\end{equation}
where
%
\begin{eqnarray}
\left(\mathcal{M}_S^2\right)_{11} 
&=& \left<\frac{\partial^2V_{\rm eff}}{\partial h_d^2}\right>,\quad
\left(\mathcal{M}_S^2\right)_{22} 
= \left<\frac{\partial^2V_{\rm eff}}{\partial h_u^2}\right>,\quad
\left(\mathcal{M}_S^2\right)_{12} = \left(\mathcal{M}_S^2\right)_{21}
= \left<\frac{\partial^2V_{\rm eff}}{\partial h_d\partial h_u}\right>,
\nonumber \\
\left(\mathcal{M}_P^2\right)_{11} 
&=& \left<\frac{\partial^2V_{\rm eff}}{\partial a_d^2}\right>,\quad
\left(\mathcal{M}_P^2\right)_{22} 
= \left<\frac{\partial^2V_{\rm eff}}{\partial a_u^2}\right>,\quad
\left(\mathcal{M}_P^2\right)_{12} = \left(\mathcal{M}_S^2\right)_{21}
= \left<\frac{\partial^2V_{\rm eff}}{\partial a_d\partial a_u}\right>,
\nonumber\\
\left(\mathcal{M}_{SP}^2\right)_{11}
&=& \left<\frac{\partial^2V_{\rm eff}}{\partial h_d\partial a_d}\right>,\quad
\left(\mathcal{M}_{SP}^2\right)_{12}
= \left<\frac{\partial^2V_{\rm eff}}{\partial h_d\partial a_u}\right>,
\nonumber\\
\left(\mathcal{M}_{SP}^2\right)_{21}
&=& \left<\frac{\partial^2V_{\rm eff}}{\partial a_d\partial h_u}\right>,\quad
\left(\mathcal{M}_{SP}^2\right)_{22}
= \left<\frac{\partial^2V_{\rm eff}}{\partial h_u\partial a_u}\right>,
              \label{eq:mass-matrix-scalar}
\end{eqnarray}
where each component is given in Appendix~B.
One can find that the pseudoscalar elements are factorized as 
%
\begin{equation}
 \MM^2_P = \left(\MM^2_P\right)_{12}
 \left(\begin{array}{cc}\tan\b & 1 \\ 1 & \cot\b\end{array}\right),
\label{m-pseudo-matrix}
\end{equation}
so that, the unphysical Goldstone mode $G^0$ can be extracted by 
%
\begin{equation}
 \left(\begin{array}{c}a_d\\ a_u\end{array}\right) =
 \left(\begin{array}{cc}\cos\b & \sin\b\\ -\sin\b & \cos\b\end{array}\right)
 \left(\begin{array}{c} G^0\\ a\end{array}\right).  \label{eq:ps-rotation}
\end{equation}
Hence, the mass-squared eigenvalues of the neutral Higgs bosons are the eigenvalues
of the matrix
%
\begin{equation}
 \MM_0^2 \equiv
 \left(\begin{array}{ccc} \left(\MM^2_S\right)_{11} & \left(\MM^2_S\right)_{12} & 
           {1\over{\cos\b}}\left(\MM^2_{SP}\right)_{12} \\
           \left(\MM^2_S\right)_{12} & \left(\MM^2_S\right)_{22} & 
           {1\over{\sin\b}}\left(\MM^2_{SP}\right)_{21} \\
           {1\over{\cos\b}}\left(\MM^2_{SP}\right)_{12} & 
		   {1\over{\sin\b}}\left(\MM^2_{SP}\right)_{21} &
           {1\over{\sin\b\cos\b}}\left(\MM^2_P\right)_{12} \end{array}\right).
                  \label{eq:mass-matrix-2}
\end{equation}
\par
In the presence of the $CP$ violation which induces the scalar-pseudoscalar mixing,
the pseudoscalar is no longer a mass eigenstate. In what follows, we use the mass of the
charged Higgs boson as an input parameter, instead of the pseudoscalar.
The mass matrix of the charged Higgs scalar has the form of
$$
 \MM^2_\pm = \expecv{{{\del^2\Veff}\over{\del\phi^+_d\del\phi^-_u}}}
 \left(\begin{array}{cc}\tan\b & 1 \\ 1 & \cot\b\end{array}\right).
$$
Similarly, we can extract the Goldstone mode so that he mass of the charged scalar 
is given by
\begin{equation}
 m_{H^\pm}^2 = {1\over{\sin\b\cos\b}}\left(\MM^2_\pm\right)_{12}.
\end{equation}
The detailed form of the mass is given in Appendix C, by which one can express
${\rm Re}(m_3^2e^{i\theta})$ in terms of $m_{H^\pm}$.\par
The effective potential at finite temperature contains temperature-dependent corrections;
\begin{equation}
 \Veff(\Phi_d,\Phi_u;T) =  \Veff(\Phi_d,\Phi_u;T=0) + \Delta_T V(\Phi_d,\Phi_u;T),
            \label{eq:Veff-T}
\end{equation}
where
\begin{eqnarray}
\Delta_T V = &&
 \frac{T^4}{2\pi^2}\left[6I_B\left(\frac{\bar{m}_W^2(\Phi)}{T^2}\right) + 
   3I_B\left(\frac{\bar{m}_Z^2(\Phi)}{T^2}\right)\right] \nonumber \\
  &&
 +6\cdot\frac{T^4}{2\pi^2}\sum_{q=t,b} \left[
 -2 I_F\left(\frac{\bar{m}_q^2(\Phi)}{T^2}\right) +
 \sum_{j=1,2}I_B\left(\frac{\bar{m}_{\tilde{t}_{q_j}}^2(\Phi)}{T^2}\right) \right].
                         \label{eq:temperature_correction}
\end{eqnarray}
Here the functions $I_B(a^2)$ and $I_F(a^2)$ are defined by
\begin{equation}
I_{B,F}(a^2)=\int^\infty_0 dx\;x^2\log\left(1\mp
  e^{-\sqrt{x^2+a^2}}\right).
\end{equation}
The function $I(a^2)$ yields $a^3$-term with negative coefficient when expanded
for $a^2\ll1$\cite{DolanJackiw}. This qualitatively explains why the EWPT becomes first order
with bosons whose field-dependent mass-squared behave as $\bar{m}^2(v)\sim v^2$ for 
small $v^2$. Because of its large Yukawa coupling, a stop with a vanishing soft mass
makes the first-order EWPT stronger.
In the following, we will not use the high-temperature expansion ($m^2/T^2\ll1$),
but numerically calculate the integrations to study the EWPT quantitatively.\par
An important constraint on the finite-temperature effective potential comes from
the requirement that the sphaleron process decouples immediately after the EWPT,
in order for baryon asymmetry produced at the EWPT not to be washed out.
If we denote the minimum of the effective potential as $(v_d, v_u, \theta)$,
the first-order EWPT is characterized by the degenerate minima at the transition
temperature $T_C$; 
$(v_C\cos\beta_C, v_C\sin\beta_C, \theta_C)$ in the broken phase region and
$(0, 0, \theta_0)$ in the symmetric phase region.
The sphaleron decoupling condition is now written as\cite{sph-decouple}
\begin{equation}
   \frac{v_C}{T_C}>1.     \label{eq:sph-decouple}
\end{equation}
The difference between $\theta_C$ and $\theta_0$ is crucial to determine total
amount of baryon asymmetry produced at the EWPT.
The profile of the bubble wall created at the first-order EWPT is derived from
the equation of motion for the gauge-Higgs system with the effective potential
at $T_C$\cite{profile}. Then the boundary conditions for the Higgs fields are
provided by $(v_C\cos\beta_C, v_C\sin\beta_C, \theta_C)$ and $(0, 0, \theta_0)$.
Since the bubble wall profile smoothly interpolates between the degenerate minima,
the phase $\theta$ varies from $\theta_0$ to $\theta_C$ at the phase boundary.
If there is no local minimum of the effective potential at $T_C$ which leads to
the transitional $CP$ violation\cite{trans-CPV}, we expect that the phase monotonously 
varies so that the baryon number generated is characterized by $\theta_C-\theta_0$.
Once we choose the phase convention such that $\theta=0$ at the zero-temperature
vacuum, both $\theta_C$ and $\theta_0$ are definitely calculated from the
effective potential in the presence of the explicit $CP$ violation in the squark
sector.
As seen from (\ref{eq:tadpole-condition}), ${\rm Im}(m_3^2)$ becomes nonzero
and $\theta_0 = -\delta$ with $\delta\equiv {\rm Arg}(m_3^2)$\footnote{%
In the second paper of \cite{profile}, the relation $\theta_0 = -\delta$ is proved 
by use of the kink ansatz for the wall profile. 
One can show that this relation generally holds, by use of the asymptotic expansion 
of the wall profile in the symmetric phase.}.
Then $\theta_C+\delta$ is another important quantity we must evaluate.
As shown in various works, when $\theta_C+\delta$ is $\mathcal{O}(1)$, 
sufficient BAU is generated by the charge transport mechanism\cite{charge-transport}. 
%
%
%
%
\section{Numerical results}
There are many parameters in the model, some of which can be fixed by
requiring the vacuum at zero temperature to be the prescribed one characterized
by $v_0=246\mbox{GeV}$, $\tanb$ and $\theta$.
In practice, we determine $m_1^2$ and $m_2^2$ by use of the tadpole condition 
(\ref{eq:tadpole-condition}). ${\rm Re}(m_3^2 e^{i\theta})$ can also be determined,
once the charged Higgs mass $m_{H^\pm}$ is given  by (\ref{eq:charged-higgs-mass}).
In order to evaluate the right-hand-sides of these equations, one must prepare
$\mu$ and the parameters in the squark sector; soft masses $m_\sq$, $m_{\st_R}$ and $m_{\sb_R}$,
trilinear couplings $A_t$ and $A_b$.
Since the $CP$ symmetry is violated by nonzero ${\rm Im}(\mu A_t)$ and ${\rm Im}(\mu A_b)$,
we take $\mu$ to be real and regard the phases of the $A$-parameters as inputs.
The EWPT is strongly first order when the soft mass of squark is very small, as noted in
the last section. One usually choose $m_\st\simeq0$ and $m_\sq=\mathcal{O}(100)\mbox{GeV}$
to avoid too large corrections to the $\rho$-parameter.
For definiteness, we take $m_{\st_R}=10\mbox{GeV}$, $m_{\sb_R}=100\mbox{GeV}$,
and several values of $m_\sq$ larger than $1\mbox{TeV}$.
As seen from the scalar-pseudoscalar mixing elements of the mass-squared matrix
(\ref{eq:mass-matrix-2}), the $CP$ violation induced in the Higgs sector is enhanced
for a larger ${\rm Im}(\mu A_t)$. In the following, we take large values for
$\mu$ and $\absv{A_t}$, but to avoid a color-and-charge-breaking vacuum, we keep
$\mu\cot\beta = A \equiv \absv{A_t} = \absv{A_b}$\footnote{In practice, we also studied
the case with $\mu\cot\beta\not=A$, but the results are not so altered as long as
the squarks do not have nonzero expectation value.}.
Thus all the input parameters are $\tanb$, $\mu$, $m_{H^\pm}$, $m_\sq$ and
$\delta_A = {\rm Arg} A_t={\rm Arg}A_b$.\par
First of all, we turn off $CP$ violation and calculate masses of the neutral Higgs bosons
for various $\tanb$, $m_{H^\pm}$, $\mu$ and $m_\sq$.  
Among parameter sets consistent with the present bounds on the lightest Higgs boson mass
$m_{H_1}$, we pick up several ones, for which the EWPT is investigated.
We numerically calculate the effective potential (\ref{eq:Veff-T}), and search for the minimum
by use of the downhill simplex algorithm.
We define $T_C$ as the temperature at which this minimum degenerates with the symmetric phase,
and evaluate $v_C$, $\tan\beta_C$ and $\theta_C$.
Next we gradually increase $\delta_A$ and examine $T_C$, $v_C$ and $\theta_C+\delta$.\par
Before showing the numerical results, we roughly describe how the spectrum, $CP$ violation
and the strength of the EWPT depend on the parameters.
If we increase $m_{H^\pm}$, which implies larger ${\rm Re}(m_3^2)$, the masses of the neutral
Higgs boson grow and scalar-pseudoscalar mixing decreases, since the diagonal elements of
 (\ref{eq:mass-matrix-2}) increases.
Then, as is well known, the EWPT becomes stronger for larger $m_{H^\pm}$.
If we decrease $\tanb$, which implies larger top Yukawa coupling $y_t$ and larger $A$
for a fixed $\mu$, the EWPT becomes stronger and Higgs-mixing is enhanced.
As for the effects of $\delta_A$, we expect that the strength of the first-order
EWPT will be weakened.
This is because nonzero $\delta_A$ modifies $v$-dependence of the mass-squared of the 
lighter stop, which is roughly proportional to $v^2$ when $m_{\st_R}=0$ and $\delta_A=0$,
as seen from (\ref{masses-in-the-vacuum-st}).\par
\begin{figure}[ht!]
 \centerline{\epsfig{file={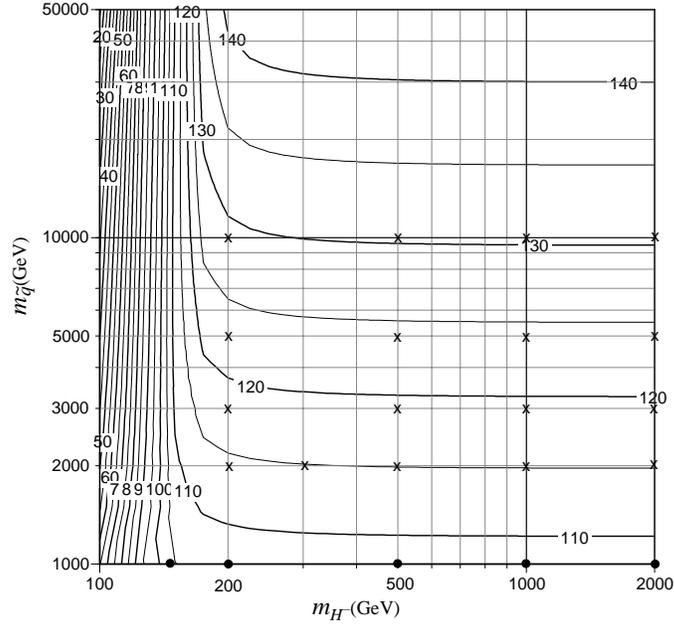}, height=90mm}}
 \caption{The lightest Higgs  mass for $\tan\b=10$ and $\mu=1500\mbox{GeV}$.
 The dot stands for parameter set for which the EWPT is strongly  first order, and
 the cross for that not satisfying (\ref{eq:sph-decouple}).}
 \label{fig:tanb=10.mu=1500}
\end{figure}

Now we present numerical results.
In the absence of $CP$ violation, the bounds on the mass of the lightest neutral Higgs 
excludes the most portion of theoretically allowed region\cite{PDG}.
For small $\tan\beta$, the lower bound is the same as the MSM, $m_h\geq 115{\rm GeV}$.
For $\tan\beta$ between $8$ and $40$, the lightest Higgs boson can be as light as
$92{\rm GeV}$. Very large $\tan\beta$-region is excluded by the fact that CDF at
Fermilab have not observed $b\bar{b}$-pair from the lightest Higgs boson.
As for the charged scalar, the bounds are satisfied with $m_{H^\pm}\geq 90{\rm GeV}$
for $\tan\beta=1 - 50$.
First of all, we turned off the $CP$ violation and calculated the masses of the Higgs 
bosons for $\tanb=5$, $10$, $20$ and $30$. For each $\tanb$, we studied the EWPT at
several points $(\mu, m_\sq)$ which are allowed by the Higgs mass bounds.
An example of contour plots of the lightest Higgs mass in $(m_{H^\pm},m_\sq)$-plane
is shown in Fig.~\ref{fig:tanb=10.mu=1500}.
There is no parameter set with $\tanb=5$ for which the EWPT is strongly first order
satisfying (\ref{eq:sph-decouple}).
For $\tanb=10$, $20$, we found several parameter sets with small $m_\sq$
for which the EWPT is strongly first order. 
The lightest Higgs boson mass is smaller than about $110\mbox{GeV}$ for such
parameter sets.\par
\begin{figure}[h!]
 \centerline{\epsfig{file={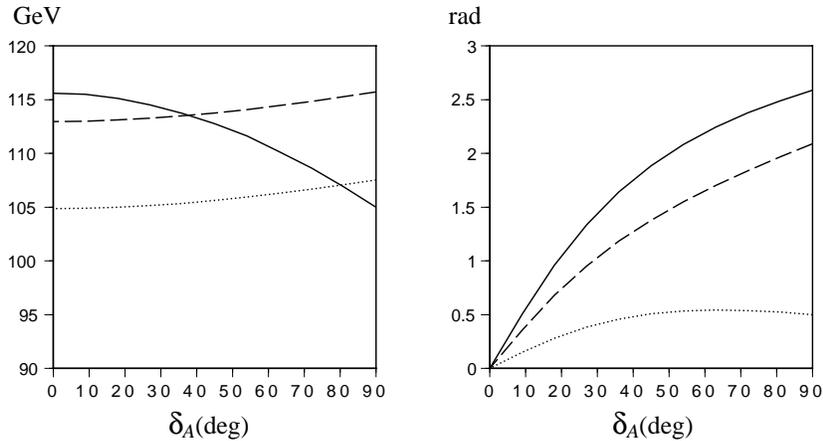}, height=60mm}}
 \caption{In the left-hand figure, the $\delta_A$-dependences of $v_C$ (solid curve), 
 $T_C$ (dashed curve) and the lightest Higgs mass $m_{H_1}$ (dotted curve) are plotted
 for $\tan\b=10$, $\mu=1500\mbox{GeV}$, $m_{H^\pm}=150\mbox{GeV}$ and $m_\sq=1\mbox{TeV}$.
 In the left-hand figure, the dashed curve stands for $\delta\equiv{\rm Arg}(m_3^2)$,
 the dotted curve for $\theta_C$ and the solid curve for $\theta_C+\delta$ for
 the same parameter set.}
 \label{fig:A-150}
\end{figure}
\begin{figure}[h!]
 \centerline{\epsfig{file={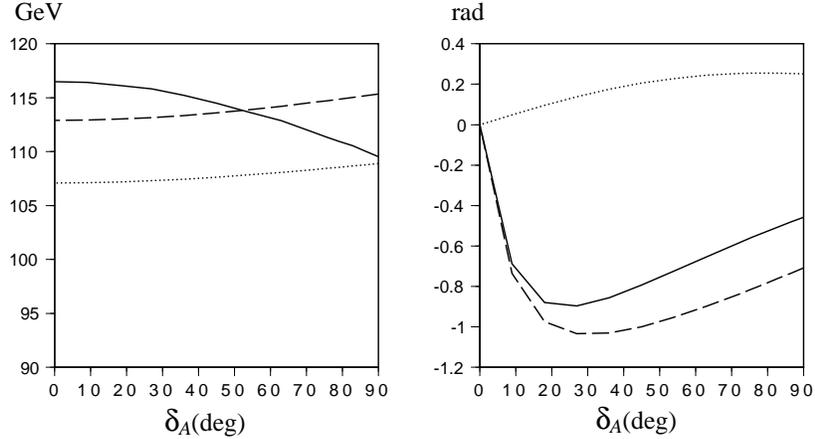}, height=60mm}}
 \caption{The same as Fig.~\ref{fig:A-150} but with $m_{H^\pm}=200\mbox{GeV}$.}
 \label{fig:A-200}
\end{figure}

Next, introducing $CP$ violating phase $\delta_A$, we studied the strength of the EWPT
and the $CP$ violation relevant to electroweak baryogenesis at several parameter sets
for which the Higgs mass bounds are satisfied and the EWPT is strongly first order
in the absence of the $CP$ violation.
In our convention, $\delta_A\simeq\pi$ is disfavored by the 
$b\rightarrow s\gamma$ constraint\cite{bsgammma}.
For example, the results are shown in Fig.~\ref{fig:A-150} for the point in 
Fig.~\ref{fig:tanb=10.mu=1500} with $m_{H^\pm}=150\mbox{GeV}$ and $m_\sq=1\mbox{TeV}$.
Although we do not show the mass of the second-lightest Higgs boson, the mixing in
the Higgs bosons becomes maximal for $\delta_A>90\mbox{deg}$.
The strength of the EWPT becomes too weak to satisfy the sphaleron decoupling condition
(\ref{eq:sph-decouple}) for $\delta_A\gtsim 40\mbox{deg}$.
The magnitude of $CP$ phase relevant to baryogenesis is sufficiently large for
$\delta_A\gtsim 10\mbox{deg}$, in spite of small Higgs mixing.
For a larger charged Higgs mass, the EWPT becomes stronger, while the Higgs mixing
becomes smaller, as expected.
The results for $m_{H^\pm}=200\mbox{GeV}$ are shown in Fig.~\ref{fig:A-200}.
The strongly first-order EWPT persists for $\delta_A\ltsim 55\mbox{deg}$,
while the magnitude of $\theta_C+\delta$ decreases.
We also examined the EWPT and the $CP$ phase for larger $m_{H^\pm}$'s and found
that the strongly first order EWPT persists for larger $\delta_A$, but 
$\absv{\theta_C+\delta}$ becomes very small. For example, the maximal value of 
$\absv{\theta_C+\delta}$ is about $0.02$ for $m_{H^\pm}=1\mbox{TeV}$ and $0.005$ for
$m_{H^\pm}=2\mbox{TeV}$.\par
\begin{figure}[h!]
 \centerline{\epsfig{file={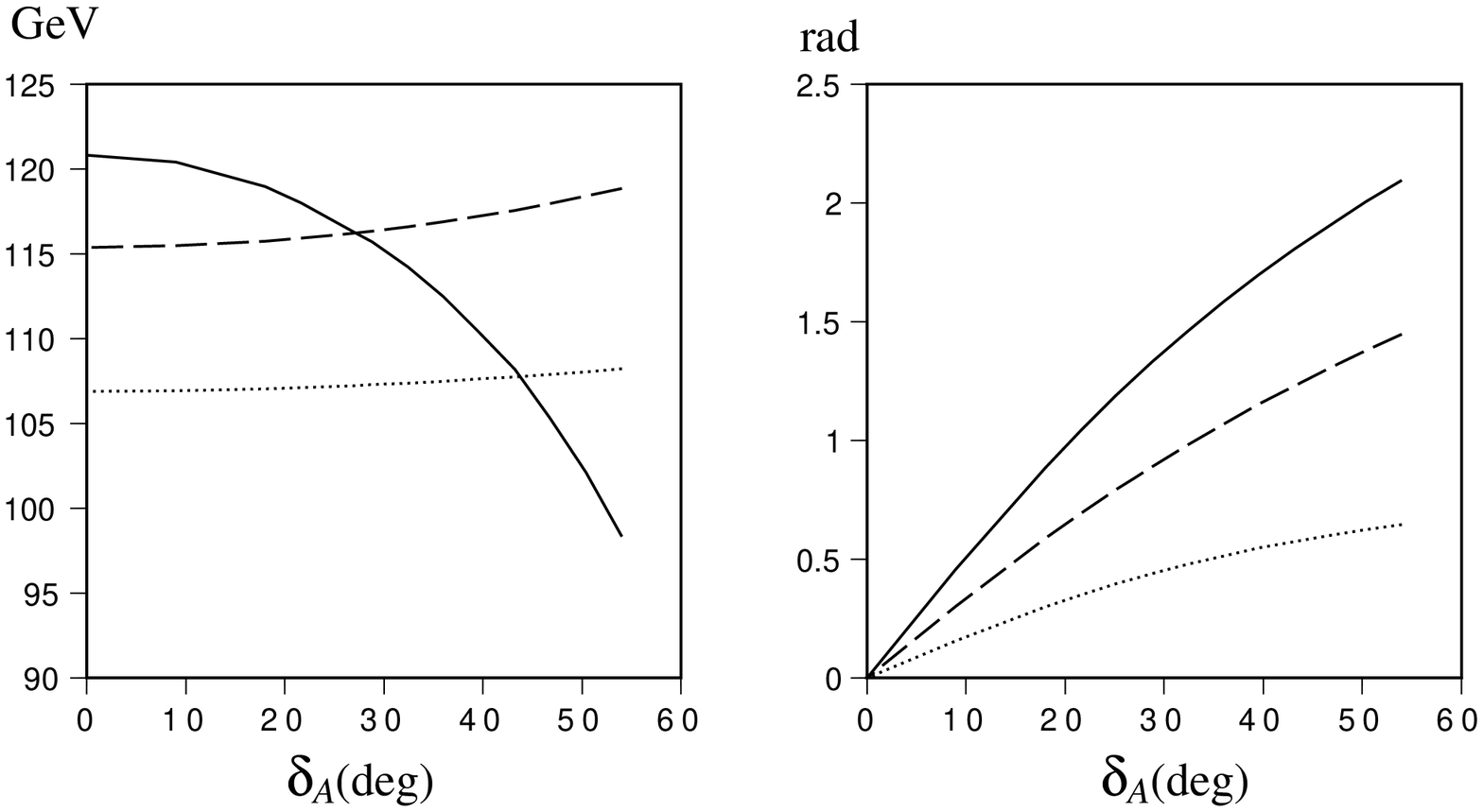}, height=60mm}}
 \caption{The same as Fig.~\ref{fig:A-200} but with $\mu=2500\mbox{GeV}$
 and $m_\sq=1100\mbox{GeV}$.}
 \label{fig:B-200}
\end{figure}
\begin{figure}[h!]
 \centerline{\epsfig{file={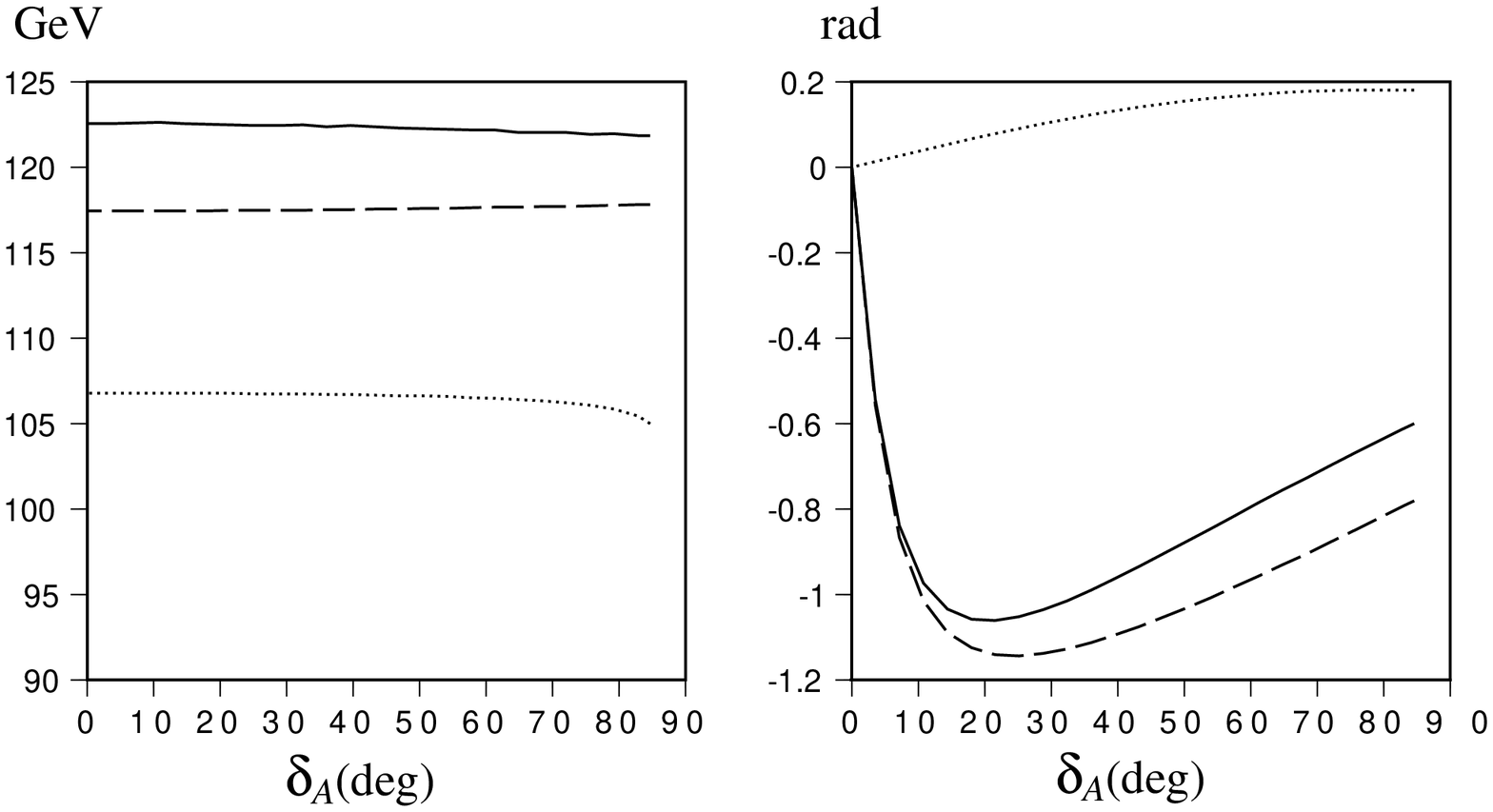}, height=60mm}}
 \caption{The same as Fig.~\ref{fig:A-200} but with $\tanb=20$, $\mu=2500\mbox{GeV}$
 and $m_\sq=1220\mbox{GeV}$.}
 \label{fig:D-200}
\end{figure}

For larger $\mu$, the effect of $\delta_A$ is expected to become stronger,
that is, the EWPT is weakened for smaller $\delta_A$.
The results for $\mu=2500\mbox{GeV}$ and $m_\sq=1100\mbox{GeV}$ are depicted
in Fig.~\ref{fig:B-200}. The strongly first-order EWPT persists for 
$\delta_A\ltsim 30\mbox{degree}$, while $\absv{\theta_C+\delta}$ is $\mathcal{O}(1)$
for $\delta_A\simeq 20\mbox{degree}$.
If we increase $m_{H^\pm}$, the EWPT gets stronger so that (\ref{eq:sph-decouple})
is satisfied up to larger $\delta_A$, but the magnitude of $CP$ violation
$\absv{\theta_C+\delta}$ decreases, as the case with smaller $\mu$.
As an example of a larger $\tanb$, we show the results for
$\tanb=20$, $\mu=2500\mbox{GeV}$ and $m_\sq=1220\mbox{GeV}$ in Fig.~\ref{fig:D-200}.
As noted above, a larger $\tanb$ implies a smaller top Yukawa coupling, so that
the effect of $CP$ violation in the stop sector decreases.
In fact, Fig.~\ref{fig:D-200} shows that the strongly first-order EWPT persists
for a larger $\delta_A$ and $\absv{\theta_C+\delta}$ is smaller than than the case with 
$\tanb=10$ (Fig.~\ref{fig:B-200}).
We also explored other parameter sets with the same $\tanb$ and observed the same
tendency as the case with $\tanb=10$.
%
%
%
\section{Conclusion}
We have studied the EWPT and $CP$ violation in the Higgs sector of the MSSM.
For parameter sets consistent with the present bounds on the masses of neutral
and charged Higgs bosons, we found strongly first-order EWPT when the lighter
stop mass is less than that of the top quark and the lightest Higgs mass
is less than about $110\mbox{GeV}$ for $8\ltsim\tanb\ltsim 30$, in the absence
of $CP$ violation. 
These results without $CP$ violation are not new.
Since the transitional $CP$ violation cannot occur for parameter sets consistent
with the updated Higgs boson mass bounds, viable $CP$ violations for electroweak
baryogenesis are among those in the complex parameters in the model.
The relative phase between $\mu$ and gaugino soft mass is essential for the scenario
in which the charginos and neutralinos play the role of charge carriers, while
the phase is strongly constrained by the neutron EDM bound.
If the masses of charginos and neutralinos are found to be as heavy as $1\mbox{TeV}$,
they cannot participate the baryogenesis, since they can hardly be excited at
the EWPT temperature.
Another source of $CP$ violation is the relative phase between $\mu$ and the squark-Higgs
trilinear coupling $A$. In particular, that phase in the third generation is free
from the nEDM constraint, and is expected to play an important role in baryogenesis.
Further, it was pointed out that the explicit $CP$ violation in the stop sector
can make the scalar-pseudoscalar mixing in the Higgs bosons very large so that 
the Higgs boson lighter than the present bound is allowed because of their
couplings to the gauge bosons and $b$-quarks\cite{Carena}.
We investigated how this phase affects the EWPT.
We found that for a larger phase, the scalar-pseudoscalar mixing increases, but
the first-order EWPT becomes weaker. The EWPT cannot continue to be strong enough 
for successful electroweak baryogenesis before the mixing reaches the maximal.
Hence the Higgs scalar lighter than the present bound, which is allowed with a large
explicit $CP$ violation, is not consistent with the strongly first-order EWPT
which is essential for electroweak baryogenesis.
We, however, found that the phase relevant for the baryogenesis can be sufficiently
large for the parameter sets consistent with the present mass bounds of the Higgs bosons.
Since baryon asymmetry produced by the electroweak baryogenesis depends on 
the phase difference of the Higgs sector between the broken and symmetric phase regions,
which are separated by the bubble wall created at the first-order EWPT.
In our phase convention, the phase is $\theta_C+\delta$. The phase can be
$\mathcal{O}(1)$ for $m_{H_1}\ltsim110\mbox{GeV}$, $m_{H^\pm}\ltsim200\mbox{GeV}$ and 
$8\ltsim\tanb\ltsim20$, and decreases for a larger $m_{H^\pm}$.\par
The calculations here are done at the one-loop level.
At finite temperatures, the naive loop expansion is not always reliable, and
the resummed perturbation with temperature-corrected masses will improve the approximation.
In particular, the infrared behavior of the Higgs scalar loop must be treated carefully.
This is the reason why the results based on perturbation in the MSM was modified
by the improved perturbation or nonperturbative lattice calculation.
As for the MSSM, the EWPT is mainly controlled by the stop loops, and the one-loop
results in Ref.\cite{KF-MSSM} are consistent with the results obtained by improved 
perturbation\cite{imp-MSSM} and nonperturbative lattice 
calculation\cite{MSSM-3Dlattice, MSSM-4Dlattice}, 
for the parameters with small $m_{\st_R}$, $\mu$ and $A$. 
We expect that the results in this work will not be altered by such improvements.\par
The parameter region of the MSSM allowed by the Higgs mass bounds is now much
narrower than that allowed theoretically.
Within that region, strongly first-order EWPT is possible only for $m_{\st_1}\ltsim m_t$
and $m_{H_1}\ltsim 110\mbox{GeV}$, which corresponds to $8\ltsim\tanb\ltsim 30$.
The explicit $CP$ violation in the stop sector can induce large $CP$ phase in the Higgs
sector which is relevant to electroweak baryogenesis. Whether this scenario works or not
depends on the spectrum of the lightest Higgs boson, charged Higgs boson and the lighter
stop. The masses of these particles will be clarified in the near future by LHC.
If the lightest Higgs boson is heavier than $110\mbox{GeV}$, the EWPT will be too weak
to make the sphaleron process out of equilibrium. Even if the EWPT is strongly first order,
the model with heavy charged Higgs boson and large $\mu$ requires another source of 
$CP$ violation than that in the Higgs sector for successful electroweak baryogenesis.
It might be the relative phases of $\mu$ and the gaugino masses, which are constrained
by nEDM experiments.
For these phase to work, the masses of the charginos and neutralinos should be
as light as the weak scale, and it will also be checked in the near future.
%
%
%
\section*{Acknowledgements}
The authors gratefully thank to A.~Kakuto and S.~Otsuki for valueble discussions.
This work is supported in part by a Grand-in-Aid for Scientific Research No.~13135222
from the Japan Society for the Promotion of Science.
%
%
\appendix
%
\section{Field-dependent masses}
In this section, we summarize the field-dependent masses of the quarks and squarks of the
third generation, and that of the gauge bosons, which appear in the definition 
of the effective potential (\ref{effective-potential-0}).
These are functions of the neutral components of the Higgs fields, while the charged
components are needed to calculate the mass of the charged Higgs boson.
The masses of the quark of the third generation are given by
\begin{equation}
\bar{m}_{t,b}^2 = \frac{1}{2}
\left[\left|y_t\right|^2\Phi_u^\dag\Phi_u + \left|y_b\right|^2\Phi_d^\dag\Phi_d
  \pm\sqrt{(\left|y_t\right|^2\Phi_u^\dag\Phi_u - \left|y_b\right|^2
    \Phi_d^\dag\Phi_d)^2 + 4\left|\Phi_d^\dag\Phi_u\right|^2}\right].
\label{quark-masses}
\end{equation}
Similarly, the top and bottom squark masses are
%
\begin{eqnarray}
 & & \bar{m}_{\tilde{t}_{1,2}}^2 = \frac{1}{2}
\left\{m_{\tilde{q}}^2 + m_{\tilde{t}_R}^2 
  + 2\left|y_t\right|^2\Phi_u^\dag\Phi_u
  + \frac{g_2^2 + g_1^2}{4}\left(\Phi_d^\dag\Phi_d - \Phi_u^\dag\Phi_u\right)
\right. \nonumber \\
 & & \left.\pm\sqrt{\left[m_{\tilde{q}}^2-m_{\tilde{t}_R}^2
        + x_t\left(\Phi_d^\dag\Phi_d - \Phi_u^\dag\Phi_u\right)\right]^2
      + 2\left|y_t\right|^2\left|\mu(v_d+h_d+ia_d)
        - A_t^*e^{-i\theta}(v_u+h_u-ia_u)\right|^2}
\right\}, \nonumber \\
\label{stop-masses} 
\end{eqnarray}
%
\begin{eqnarray}
 & & \bar{m}_{\tilde{b}_{1,2}}^2 = \frac{1}{2}
\left\{m_{\tilde{q}}^2 + m_{\tilde{b}_R}^2 
  + 2\left|y_b\right|^2\Phi_d^\dag\Phi_d
  - \frac{g_2^2 + g_1^2}{4}\left(\Phi_d^\dag\Phi_d - \Phi_u^\dag\Phi_u\right)
\right. \nonumber \\
 & & \left.\pm\sqrt{\left[m_{\tilde{q}}^2-m_{\tilde{b}_R}^2
        + x_b\left(\Phi_d^\dag\Phi_d - \Phi_u^\dag\Phi_u\right)\right]^2
      + 2\left|y_b\right|^2\left|\mu e^{i\theta}(v_u+h_u+ia_u)
        - A_b^*(v_d+h_d-ia_d)\right|^2}
\right\}, \nonumber \\
\label{sbottom-masses}
\end{eqnarray}
where
%
\begin{equation}
x_t\equiv\frac{1}{4}\left(g_2^2-\frac{5}{3}g_1^2\right),\quad
x_b\equiv-\frac{1}{4}\left(g_2^2-\frac{1}{3}g_1^2\right).
\label{xt-xb}
\end{equation}
The gauge bosons masses are
%
\begin{eqnarray}
\bar{m}_Z^2 &=& \frac{1}{4}(g_2^2+g_1^2)\left[(v_d+h_d)^2+a_d^2
  +(v_u+h_u)^2+a_u^2\right]\label{mzbar}                                    \label{mzbar}\\
\bar{m}_W^2 &=& \frac{1}{4}g_2^2\left[(v_d+h_d)^2+a_d^2+(v_u+h_u)^2\right]. \label{mwbar}
\end{eqnarray}
The masses evaluated at the zero-temperature vacuum are given as follows:
%
\begin{eqnarray}
 m^2_t &=& \expecv{\barm^2_t}=\half\absv{y_t}^2v_u^2,\qquad
 m^2_b  =  \expecv{\barm^2_b}=\half\absv{y_b}^2v_d^2,
\label{masses-in-the-vacuum-tb}   \\
 m_{\st_{1,2}}^2 &=& \expecv{\barm_{\st_{1,2}}^2}  \nonumber\\
 &=&
 {{m_\sq^2+m_{\st_R}^2}\over2}+m_t^2+ {1\over4}m_Z^2\cos(2\b)   \nonumber\\
 &&\qquad \pm
 \half\sqrt{\left[m_\sq^2-m_{\st_R}^2+{{x_t}\over2}v_0^2\cos(2\b)\right]^2
           +4m_t^2\absv{\mu\cot\b-A_t^*e^{-i\th}}^2}
\label{masses-in-the-vacuum-st}      \\
 m_{\sb_{1,2}}^2 &=& \expecv{\barm_{\sb_{1,2}}^2}  \nonumber\\
 &=&
 {{m_\sq^2+m_{\sb_R}^2}\over2}+m_b^2- {1\over4}m_Z^2\cos(2\b)   \nonumber\\
 &&\qquad \pm
 \half\sqrt{\left[m_\sq^2-m_{\sb_R}^2+{{x_b}\over2}v_0^2\cos(2\b)\right]^2
           +4m_b^2\absv{\mu e^{i\th}\tan\b-A_b^*}^2},
\label{masses-in-the-vacuum-sb}  \\
 m_Z^2 &=& {1\over4}(g_2^2+g_1^2)v_0^2,\qquad
 m_W^2  =  {1\over4}g_2^2v_0^2,
\label{masses-in-the-vacuum-ZW}
\end{eqnarray}
where $v_0^2\equiv v_d^2+v_u^2$, $\tan\b=v_u/v_d$.
%
\section{Mass matrix of the neutral Higgs boson}
The calculation of the elements of the mass matrix is presented in Ref.\cite{Carena}.
We included the gauge-boson contributions to their calculation, so that
we record their explicit forms.
For later convenience, we introduce the following quantities:
%
\begin{eqnarray}
 && M_t^2 = m_\sq^2-m_{\st_R}^2+{{x_t}\over2}v_0^2\cos(2\b), \qquad
    M_b^2 = m_\sq^2-m_{\sb_R}^2+{{x_b}\over2}v_0^2\cos(2\b),  \\
 && \Delta m_\sq^2 = m_{\sq_1}^2 - m_{\sq_2}^2,\qquad (q=t,b)  \\
 && R_q = {\rm Re}\left(\mu A_q e^{i\th}\right),\quad
    I_q = {\rm Im}\left(\mu A_q e^{i\th}\right),\qquad (q=t,b)  \\
 && P_t = \absv{\mu}^2 - R_t \tan\b, \qquad
    Q_t = \absv{A_t}^2 - R_t \cot\b,    \\
 && P_b = \absv{\mu}^2 - R_b \cot\b, \qquad
    Q_b = \absv{A_b}^2 - R_b \tan\b.
\end{eqnarray}
The elements of the mass matrix in the scalar sector are given by
%
\begin{eqnarray}
 \left(\MM^2_S\right)_{11}  \!\!\!&=&\!\!\!
 {\rm Re}(m_3^2e^{i\th})\tan\b + m_Z^2\cos^2\b  \nonumber\\
 &&\!\!\!+
 {{N_C}\over{16\pi^2}}\Biggl\{
 \half m_Z^2\cos^2\b \left(
  {{x_tM_t^2+2\absv{y_t}^2P_t}\over{\Delta m_\st^2}}\log{{m_{\st_1}^2}\over{m_{\st_2}^2}}
 -{{x_bM_b^2+2\absv{y_b}^2Q_b}\over{\Delta m_\sb^2}}\log{{m_{\sb_1}^2}\over{m_{\sb_2}^2}}
                  \right)                \nonumber\\
 &&\!\!\!+            
 2m_b^2\left( {{x_bM_b^2+2\absv{y_b}^2Q_b}\over{\Delta m_\sb^2}}
               \log{{m_{\sb_1}^2}\over{m_{\sb_2}^2}}
              -2\absv{y_b}^2 \log{{m_b^2}\over{M^2}} \right)          \nonumber\\
 &&\!\!\!+            
  v_d^2\left({{g_2^2+g_1^2}\over8}\right)^2
       \left(\log{{m_{\st_1}^2}\over{M^2}}+\log{{m_{\st_2}^2}\over{M^2}}\right)
 +v_d^2\left(\absv{y_b}^2-{{g_2^2+g_1^2}\over8}\right)^2
       \left(\log{{m_{\sb_1}^2}\over{M^2}}+\log{{m_{\sb_2}^2}\over{M^2}}\right) \nonumber\\
 &&\!\!\!+            
 {{x_t^2v_d^2+2\absv{y_t}^2R_t\tan\b}\over{2\Delta m_\st^2}}
 \left(m_{\st_1}^2\log{{m_{\st_1}^2}\over{M^2}}-m_{\st_2}^2\log{{m_{\st_2}^2}\over{M^2}}
       - \Delta m_\st^2 \right)   \nonumber\\
 &&\!\!\!+            
 {{x_b^2v_d^2+2\absv{y_b}^2R_b\tan\b}\over{2\Delta m_\sb^2}}
 \left(m_{\sb_1}^2\log{{m_{\sb_1}^2}\over{M^2}}-m_{\sb_2}^2\log{{m_{\sb_2}^2}\over{M^2}}
       - \Delta m_\sb^2 \right)   \nonumber\\
 &&\!\!\!+            
 v_d^2{{(x_tM_t^2+2\absv{y_t}^2P_t)^2}\over{2(\Delta m_\st^2)^2}}
  \left(1 - {{m_{\st_1}^2+m_{\st_2}^2}\over{2\Delta m_\st^2}}
             \log{{m_{\st_1}^2}\over{m_{\st_2}^2}}\right)   \nonumber\\
 &&\!\!\!+            
 v_d^2{{(x_bM_b^2+2\absv{y_b}^2Q_b)^2}\over{2(\Delta m_\sb^2)^2}}
  \left(1 - {{m_{\sb_1}^2+m_{\sb_2}^2}\over{2\Delta m_\sb^2}}
              \log{{m_{\sb_1}^2}\over{m_{\sb_2}^2}}\right) \Biggr\}  \nonumber\\
 &&\!\!\!+
 {{3}\over{128\pi^2}}v_d^2
  \left\{(g_2^2+g_1^2)^2\log{{m_Z^2}\over{M^2}}
    + 2g_2^4\log{{m_W^2}\over{M^2}}\right\},
\label{m-scalar-11}
\end{eqnarray}
%
\begin{eqnarray}
 \left(\MM^2_S\right)_{22}  \!\!\!&=&\!\!\!
 {\rm Re}(m_3^2e^{i\th})\cot\b + m_Z^2\sin^2\b  \nonumber\\
 &&\!\!\!+
 {{N_C}\over{16\pi^2}}\Biggl\{
 \half m_Z^2\sin^2\b \left(
 -{{-x_tM_t^2+2\absv{y_t}^2Q_t}\over{\Delta m_\st^2}}\log{{m_{\st_1}^2}\over{m_{\st_2}^2}}
 +{{-x_bM_b^2+2\absv{y_b}^2P_b}\over{\Delta m_\sb^2}}\log{{m_{\sb_1}^2}\over{m_{\sb_2}^2}}
                  \right)                \nonumber\\
 &&\!\!\!+            
 2m_t^2\left(
  {{-x_tM_t^2+2\absv{y_t}^2Q_t}\over{\Delta m_\st^2}}\log{{m_{\st_1}^2}\over{m_{\st_2}^2}}
  -2\absv{y_t}^2 \log{{m_t^2}\over{M^2}} \right)          \nonumber\\
 &&\!\!\!+            
  v_u^2\left(\absv{y_t}^2-{{g_2^2+g_1^2}\over8}\right)^2
       \left(\log{{m_{\st_1}^2}\over{M^2}}+\log{{m_{\st_2}^2}\over{M^2}}\right)
 +v_u^2\left({{g_2^2+g_1^2}\over8}\right)^2
       \left(\log{{m_{\sb_1}^2}\over{M^2}}+\log{{m_{\sb_2}^2}\over{M^2}}\right) \nonumber\\
 &&\!\!\!+            
 {{x_t^2v_u^2+2\absv{y_t}^2R_t\cot\b}\over{2\Delta m_\st^2}}
 \left(m_{\st_1}^2\log{{m_{\st_1}^2}\over{M^2}}-m_{\st_2}^2\log{{m_{\st_2}^2}\over{M^2}}
       - \Delta m_\st^2 \right)   \nonumber\\
 &&\!\!\!+            
 {{x_b^2v_u^2+2\absv{y_b}^2R_b\cot\b}\over{2\Delta m_\sb^2}}
 \left(m_{\sb_1}^2\log{{m_{\sb_1}^2}\over{M^2}}-m_{\sb_2}^2\log{{m_{\sb_2}^2}\over{M^2}}
       - \Delta m_\sb^2 \right)   \nonumber\\
 &&\!\!\!+            
 v_u^2{{(-x_tM_t^2+2\absv{y_t}^2Q_t)^2}\over{2(\Delta m_\st^2)^2}}
  \left(1 - {{m_{\st_1}^2+m_{\st_2}^2}\over{2\Delta m_\st^2}}
             \log{{m_{\st_1}^2}\over{m_{\st_2}^2}}\right)   \nonumber\\
 &&\!\!\!+            
 v_u^2{{(-x_bM_b^2+2\absv{y_b}^2P_b)^2}\over{2(\Delta m_\sb^2)^2}}
  \left(1 - {{m_{\sb_1}^2+m_{\sb_2}^2}\over{2\Delta m_\sb^2}}
              \log{{m_{\sb_1}^2}\over{m_{\sb_2}^2}}\right) \Biggr\}  \nonumber\\
 &&\!\!\!+
 {{3}\over{128\pi^2}}v_u^2
  \left\{(g_2^2+g_1^2)^2\log{{m_Z^2}\over{M^2}}
    + 2g_2^4\log{{m_W^2}\over{M^2}}\right\},     
\label{m-scalar-22}
\end{eqnarray}
%
\begin{eqnarray}
 \left(\MM^2_S\right)_{12}  \!\!\!&=&\!\!\!
 - {\rm Re}(m_3^2e^{i\th}) - m_Z^2\sin\b\cos\b  \nonumber\\
 &&\!\!\!+
 {{N_C}\over{16\pi^2}}\Biggl\{
 \half m_Z^2\sin\b\cos\b \left(
  {{-x_tM_t^2+\absv{y_t}^2(Q_t-P_t)}\over{\Delta m_\st^2}}\log{{m_{\st_1}^2}\over{m_{\st_2}^2}}
                         \right.          \nonumber\\
 &&\qquad\qquad\qquad\qquad\qquad\qquad\left.
 +{{x_bM_b^2+\absv{y_b}^2(Q_b-P_b)}\over{\Delta m_\sb^2}}\log{{m_{\sb_1}^2}\over{m_{\sb_2}^2}}
                  \right)                \nonumber\\
 &&\!\!\!+
  m_t^2\cot\b{{x_tM_t^2+2\absv{y_t}^2P_t}\over{\Delta m_\st^2}}
             \log{{m_{\st_1}^2}\over{m_{\st_2}^2}}
 +m_b^2\tan\b{{-x_bM_b^2+2\absv{y_b}^2P_b}\over{\Delta m_\sb^2}}
             \log{{m_{\sb_1}^2}\over{m_{\sb_2}^2}}     \nonumber\\
 &&\!\!\!+
 \half m_Z^2\sin\b\cos\b \left[
  \left(\absv{y_t}^2-{{g_2^2+g_1^2}\over8}\right)
  \left(\log{{m_{\st_1}^2}\over{M^2}}+\log{{m_{\st_2}^2}\over{M^2}}\right)
                  \right.          \nonumber\\
 &&\qquad\qquad\qquad\qquad\qquad\qquad\left. +
  \left(\absv{y_b}^2-{{g_2^2+g_1^2}\over8}\right)
  \left(\log{{m_{\sb_1}^2}\over{M^2}}+\log{{m_{\sb_2}^2}\over{M^2}}\right) \right] \nonumber\\
 &&\!\!\! -
 {{x_t^2v_dv_u+2\absv{y_t}^2R_t}\over{2\Delta m_\st^2}}
 \left(m_{\st_1}^2\log{{m_{\st_1}^2}\over{M^2}}-m_{\st_2}^2\log{{m_{\st_2}^2}\over{M^2}}
       - \Delta m_\st^2 \right)   \nonumber\\
 &&\!\!\! -
 {{x_b^2v_dv_u+2\absv{y_b}^2R_b}\over{2\Delta m_\sb^2}}
 \left(m_{\sb_1}^2\log{{m_{\sb_1}^2}\over{M^2}}-m_{\sb_2}^2\log{{m_{\sb_2}^2}\over{M^2}}
       - \Delta m_\sb^2 \right)   \nonumber\\
 &&\!\!\!+
 v_dv_u{{(x_tM_t^2+2\absv{y_t}^2P_t)(-x_tM_t^2+2\absv{y_t}^2Q_t)}\over{2(\Delta m_\st^2)^2}}
  \left(1 - {{m_{\st_1}^2+m_{\st_2}^2}\over{2\Delta m_\st^2}}
             \log{{m_{\st_1}^2}\over{m_{\st_2}^2}}\right)   \nonumber\\
 &&\!\!\!+
 v_dv_u{{(x_bM_b^2+2\absv{y_b}^2Q_b)(-x_bM_b^2+2\absv{y_b}^2P_b)}\over{2(\Delta m_\sb^2)^2}}
  \left(1 - {{m_{\sb_1}^2+m_{\sb_2}^2}\over{2\Delta m_\sb^2}}
              \log{{m_{\sb_1}^2}\over{m_{\sb_2}^2}}\right) \Biggr\}  \nonumber \\
 &&\!\!\!+
 {{3}\over{128\pi^2}}v_dv_u
  \left\{(g_2^2+g_1^2)^2\log{{m_Z^2}\over{M^2}}
    + 2g_2^4\log{{m_W^2}\over{M^2}}\right\}.
\label{m-scalar-12}
\end{eqnarray}
The matrix elements in the pseudoscalar sector are
%
\begin{eqnarray}
\label{m-pseudo-11}
 \left(\MM^2_P\right)_{11} \!\!\!&=&\!\!\! \left(\MM^2_P\right)_{12}\tan\b,   \\
\label{m-pseudo-22}
 \left(\MM^2_P\right)_{22} \!\!\!&=&\!\!\! \left(\MM^2_P\right)_{12}\cot\b,   \\
 \left(\MM^2_P\right)_{12} \!\!\!&=&\!\!\! {\rm Re}(m_3^2e^{i\th}) +
 {{N_C}\over{16\pi^2}}\Biggl\{
 {{\absv{y_t}^2}\over{\Delta m_\st^2}}\left[
  R_t \left(m_{\st_1}^2\log{{m_{\st_1}^2}\over{M^2}}-m_{\st_2}^2\log{{m_{\st_2}^2}\over{M^2}}
              - \Delta m_\st^2 \right)    \right.   \nonumber\\
 &&\qquad\qquad\qquad\qquad\left.+
  {{4m_t^2\cot\b}\over{\Delta m_\st^2}}I_t^2
  \left(1 - {{m_{\st_1}^2+m_{\st_2}^2}\over{2\Delta m_\st^2}}
             \log{{m_{\st_1}^2}\over{m_{\st_2}^2}}\right) \right]     \nonumber\\
 &&\qquad\qquad\qquad\quad+
 {{\absv{y_b}^2}\over{\Delta m_\sb^2}}\left[
  R_b \left(m_{\sb_1}^2\log{{m_{\sb_1}^2}\over{M^2}}-m_{\sb_2}^2\log{{m_{\sb_2}^2}\over{M^2}}
             - \Delta m_\sb^2 \right)    \right.   \nonumber\\
 &&\qquad\qquad\qquad\qquad\left.+
  {{4m_b^2\tan\b}\over{\Delta m_\sb^2}}I_b^2
  \left(1 - {{m_{\sb_1}^2+m_{\sb_2}^2}\over{2\Delta m_\sb^2}}
              \log{{m_{\sb_1}^2}\over{m_{\sb_2}^2}}\right) \right] \Biggr\}.
\label{m-pseudo-12}
\end{eqnarray}
The scalar-pseudoscalar mixing elements are given by
%
\begin{eqnarray}
\label{m-scalar-pseudo-11}
 \left(\MM^2_{SP}\right)_{11} \!\!\!&=&\!\!\! \left(\MM^2_{SP}\right)_{12}\tan\b,   \\
\label{m-scalar-pseudo-22}
 \left(\MM^2_{SP}\right)_{22} \!\!\!&=&\!\!\! \left(\MM^2_{SP}\right)_{21}\cot\b,   \\
 \left(\MM^2_{SP}\right)_{12} \!\!\!&=&\!\!\!
 {{N_C}\over{8\pi^2}}\Biggl\{
  {{m_t^2I_t\cot^2\b}\over{\Delta m_\st^2}}\left[
   {{g_2^2+g_1^2}\over8}\log{{m_{\st_1}^2}\over{m_{\st_2}^2}}
  +{{x_tM_t^2+2\absv{y_t}^2P_t}\over{\Delta m_\st^2}}
   \left(1 - {{m_{\st_1}^2+m_{\st_2}^2}\over{2\Delta m_\st^2}}
             \log{{m_{\st_1}^2}\over{m_{\st_2}^2}}\right) \right]   \nonumber\\[3mm]
 &&\!\!\!+
  {{m_b^2I_b}\over{\Delta m_\sb^2}}\left[
   \left(\absv{y_b}^2-{{g_2^2+g_1^2}\over8}\right)\log{{m_{\sb_1}^2}\over{m_{\sb_2}^2}}
  +{{x_bM_b^2+2\absv{y_b}^2Q_b}\over{\Delta m_\sb^2}}
   \left(1 - {{m_{\sb_1}^2+m_{\sb_2}^2}\over{2\Delta m_\sb^2}}
              \log{{m_{\sb_1}^2}\over{m_{\sb_2}^2}}\right)  \right] \Biggr\}  \nonumber\\
\label{m-scalar-pseudo-12}
 &&   \\
 \left(\MM^2_{SP}\right)_{21} \!\!\!&=&\!\!\!
 {{N_C}\over{8\pi^2}}\Biggl\{
 {{m_t^2I_t}\over{\Delta m_\st^2}}\left[
  \left(\absv{y_t}^2-{{g_2^2+g_1^2}\over8}\right)\log{{m_{\st_1}^2}\over{m_{\st_2}^2}}
 +{{-x_tM_t^2+2\absv{y_t}^2Q_t}\over{\Delta m_\st^2}}
   \left(1 - {{m_{\st_1}^2+m_{\st_2}^2}\over{2\Delta m_\st^2}}
             \log{{m_{\st_1}^2}\over{m_{\st_2}^2}}\right) \right]   \nonumber\\[3mm]
 &&\!\!\!+
  {{m_b^2I_b\tan^2\b}\over{\Delta m_\sb^2}}\left[
   {{g_2^2+g_1^2}\over8}\log{{m_{\sb_1}^2}\over{m_{\sb_2}^2}}
  +{{-x_bM_b^2+2\absv{y_b}^2P_b}\over{\Delta m_\sb^2}}
   \left(1 - {{m_{\sb_1}^2+m_{\sb_2}^2}\over{2\Delta m_\sb^2}}
              \log{{m_{\sb_1}^2}\over{m_{\sb_2}^2}}\right)  \right] \Biggr\}, \nonumber\\
\label{m-scalar-pseudo-21}
\end{eqnarray}
which are all composed of the terms proportional to ${\rm Im}(\mu A_t)/\Delta m_\st^2$ 
or ${\rm Im}(\mu A_b)/\Delta m_\sb^2$.
%
%
\section{Charged Higgs mass}
The calculation of the charged Higgs mass is a tedious task, and the method is
described in Appendix of Ref.\cite{Carena}.
We present the result, which contain the contribution from the gauge bosons:
%
\begin{eqnarray}
 m_{H^\pm}^2 \!\!\!&=&\!\!\!
 {1\over{\sin\b\cos\b}}{\rm Re}(m_3^2e^{i\th}) + m_W^2  \nonumber\\
 &+&\!\!\!
 {{N_C}\over{16\pi^2\sin\b\cos\b}}\Biggl\{
  {1\over{\Delta m_\st^2}}\left[
   \left( {{f(m_{\st_1}^2)}\over{(m_{\st_1}^2-m_{\sb_1}^2)(m_{\st_1}^2-m_{\sb_2}^2)}}
           + \absv{y_t}^2 R_t \right) m_{\st_1}^2\left(\log{{m_{\st_1}^2}\over{M^2}}-1\right)
                   \right.\nonumber\\
 &&\qquad\qquad\qquad\qquad\qquad\qquad\left.
   - (\st_1\rightarrow\st_2) \right]   \nonumber\\
 &&\qquad
 +
  {1\over{\Delta m_\sb^2}}\left[
   \left( {{f(m_{\sb_1}^2)}\over{(m_{\sb_1}^2-m_{\st_1}^2)(m_{\sb_1}^2-m_{\st_2}^2)}}
           + \absv{y_b}^2 R_b \right) m_{\sb_1}^2\left(\log{{m_{\sb_1}^2}\over{M^2}}-1\right)
   - (\sb_1\rightarrow\sb_2) \right]   \nonumber\\
 &&\qquad
 -{{2\absv{y_ty_b}m_tm_b}\over{m_t^2-m_b^2}}
  \left[ m_t^2\left(\log{{m_t^2}\over{M^2}}-1\right) 
       - m_b^2\left(\log{{m_t^2}\over{M^2}}-1\right) \right] \Biggr\}  \nonumber\\
 &-&\!\!\!
 {3\over{16\pi^2}}(g_2^2+g_1^2){{g_2^2}\over{g_1^2}}m_Z^2
 \left(\log{{m_Z^2}\over{M^2}}-1\right),               \label{eq:charged-higgs-mass}
\end{eqnarray}
where
\begin{eqnarray}
 \lefteqn{
f(m_{\sq_k}^2)} \nonumber\\
 \!\!\!&=&\!\!\!
 \half\absv{y_ty_b}^2 v_uv_d\left[ 2m_{\sq_k}^4 -{\rm Tr}(\MM_\st^2+\MM_\sb^2)m_{\sq_k}^2
    + \left(\MM_\st^2\right)_{11}\left(\MM_\sb^2\right)_{11} 
	+ \left(\MM_\st^2\right)_{22}\left(\MM_\sb^2\right)_{22} \right]   \nonumber\\
 &&
 -{1\over4}g_2^2v_uv_d\left(\absv{y_t}^2+\absv{y_b}^2-{{g_2^2}\over2}\right)
  (m_{\sq_k}^2 - \left(\MM_\st^2\right)_{22})
  (m_{\sq_k}^2 - \left(\MM_\sb^2\right)_{22})     \nonumber\\
 &&
 +\half\absv{\mu}^2v_uv_d\Biggl[
  -(\absv{y_t}^2+\absv{y_b}^2)\left(\absv{y_t}^2+\absv{y_b}^2-{{g_2^2}\over2}\right)m_{\sq_k}^2
       \nonumber\\
 &&\qquad\qquad\qquad
  +\absv{y_t}^2\left(\absv{y_t}^2-{{g_2^2}\over2}\right)\left(\MM_\sb^2\right)_{22}
  +\absv{y_b}^2\left(\absv{y_b}^2-{{g_2^2}\over2}\right)\left(\MM_\st^2\right)_{22}   \nonumber\\
 &&\qquad\qquad\qquad
  +\absv{y_ty_b}^2\left(\absv{\mu}^2 +2m_\sq^2 +2m_t^2 +2m_b^2 -m_W^2
                       -{{g_1^2}\over{12}}v_0^2\cos(2\b) \right) \Biggr]  \nonumber\\
 &&
 +\half\absv{y_tA_t}^2v_uv_d\left(\absv{y_b}^2-{{g_2^2}\over2}\right)
                          (m_{\sq_k}^2 - \left(\MM_\sb^2\right)_{22})
 +\half\absv{y_bA_b}^2v_uv_d\left(\absv{y_t}^2-{{g_2^2}\over2}\right)
                          (m_{\sq_k}^2 - \left(\MM_\st^2\right)_{22})    \nonumber\\
 &&
 +\half\absv{y_ty_b A_tA_b}^2v_uv_d - \absv{y_ty_b}^2v_uv_d(R_tR_b + I_tI_b)  \nonumber\\
 &&
 -2m_t^2m_b^2\left[\left(\absv{y_b}^2-{{g_2^2}\over2}\right) R_t
                 + \left(\absv{y_t}^2-{{g_2^2}\over2}\right) R_b \right]  \nonumber\\
 &&
 +\half\absv{y_ty_b}^2v_uv_d {\rm Re}(A_tA_b^*)
  \left[2m_{\sq_k}^2 -2m_\sq^2 - m_W^2 +{{g_1^2}\over{12}}v_0^2\cos(2\b) \right].
  \label{definition-of-f}
\end{eqnarray}
%
%
%

%
%
%
%
%
\end{document}